\documentstyle[aps,prb,epsf]{revtex}
\preprint{}
\begin{document}
\title{Rotating Bose gas with hard-core repulsion in a
quasi-2D harmonic trap: vortices in BEC}
\author{M. A. H. Ahsan and N. Kumar}
 \address{Raman Research Institute, Sir C. V. Raman Avenue, 
 Bangalore 560080, India.}
\maketitle
\begin{abstract}
We consider a gas of 
N$\left(=6, 10, 15\right)$ 
Bose particles with  
 hard-core repulsion, contained in a quasi-2D harmonic
trap and subjected to an overall angular velocity $\Omega $ about the
z-axis. 
Exact diagonalization of the $n\times n$ many-body 
Hamiltonian matrix in given subspaces of the total (quantized) angular 
momentum L$_{z}$, with $n\sim 10^{5}$(e.g. for L$_{z}$=N=15, 
  $n$ =240782) was carried 
 out using Davidson's algorithm. 
 The many-body variational ground state wavefunction, as also
the corresponding energy and the  
 reduced one-particle density-matrix
$\rho\left({\bf r},{\bf r}^{\prime }\right)=
\sum_{\mu }\lambda_{\mu }\chi_{\mu }^{\ast }\left({\bf r}\right)
\chi_{\mu }\left({\bf r}^{\prime }\right)$ 
 were calculated.
 With the usual identification of $\Omega $
as the Lagrange multiplier associated with L$_{z}$ for a rotating system, the
L$_{z}-\Omega $
 phase diagram $\left(\mbox{or the stability line}\right)$
 was determined that gave a number of critical angular velocities
$\Omega_{{\bf c}i},\ i=1,2,3,\cdots ,$ at which the ground
state angular momentum and the associated condensate fraction,
 given by the largest eigenvalue of 
 the reduced one-particle density-matrix, undergo abrupt jumps.
 A number of (total) angular momentum states were found to be stable
 at successively 
 higher critical angular velocities $\Omega_{{\bf c}i},\ i=1,2,3,\cdots $
 for a given N. 
All the states in the regime $\mbox{N}>\mbox{L}_{z}>0$ are metastable. 
 For L$_{z}>$N, the L$_{z}$ values for the stable ground states 
 generally increased with the increasing critical angular velocities
 $\Omega_{{\bf c}i}$, and the condensate was strongly depleted.
The critical $\Omega_{{\bf c}i} $ values, however, decreased with increasing 
interaction strength as well as the particle number, 
 and were systematically greater than the non-variational Yrast-state   
 values for the single vortex state with L$_{z}$=N. 
We have also observed that the condensate fraction
 for the single vortex state 
 (as also for the higher vortex states) did not change significantly  
 even as the 2-body interaction strength was varied over 
 several$\left(\sim 4\right)$ orders of magnitude in the moderately to the
 weakly interacting regime.
\end{abstract}
\section{INTRODUCTION}
The experimental realization of Bose-Einstein Condensation(BEC)
in dilute vapors of ultra-cold(nanokelvin) alkali 
atoms\cite{amw95,dma95,bst95}, and other systems \cite{klp98}, 
 trapped in harmonic
 potential wells has qualitatively extended the domain of occurrence
 of the quantum fluids\cite{bp96,dgp99,efs99}.  
 Unlike their dense and strongly interacting
 homogeneous (bulk) counterparts, e.g. liquid $^{4}$He, these
mesoscopic gaseous systems are dilute, weakly or moderately interacting 
and inhomogeneous$-$with controllable density, effective dimensionality, 
 and tunable atom-atom interactions of either sign\cite{fsbr}. 
 Further, the creation \cite{mcw00,mah99} of the vortex 
 states with quantized circulation in externally rotated traps, as also
 the direct observation of phase coherence effects\cite{atm97},  
 has clearly revealed the phase rigidity 
 characteristic of superfluidity associated with the BEC.  
 The BEC in a weakly and repulsively interacting dilute Bose gas 
({\em i.e.} with the  
 s-wave scattering length $a_{sc}\ll $ the mean inter-atomic 
 spacing and with the number of atoms in the condensate N$\gg$ 1) 
 has often been described macroscopically through the Gross-Pitaevskii 
equation based on the condensate amplitude as a slowly varying 
 order-parameter \cite{dgp99}. Microscopic treatments going beyond the
 meanfield approximation and based on the many-body
variational wavefunctions also exist, but involve heavy
computations\cite{gfb99,wg00} even for modest size of the system  
$\left(\mbox{N}\sim 10-50\right)$. Besides, these
 many-body calculations have involved only the ``Lowest Landau
Level(LLL)'' single-particle orbitals, with only the positive 
sign(with respect to the trap angular velocity $\Omega \hat{e}_{z}$) of the
 single-particle angular momentum quantum number $m$.
The inter-atomic repulsion is known to qualitatively
change the ground state properties of a system of weakly interacting
 neutral Bose gas (confined in a rotating harmonic trap) at T=0 in  
 two distinct ways \cite{dgp99}. First, it leads to a depletion of the 
condensate  
fraction, which is equal to 1 for an ideal(non-interacting) 
 Bose gas. Second, it gives a phase rigidity, or stiffness, to
the ground state many-body wavefunction. This is responsible for 
the superfluid flow that manifests in the successive appearance of  
 vortices with higher  
 quantized circulations beyond certain critical angular velocities
 $\Omega_{ci}$ of the rotating trap.
 Both these aspects are well described by the one-particle
  reduced density-matrix $\rho_{1}\left({\bf r}, {\bf r}^{\prime }\right)$ 
obtained from the N-body ground-state wavefunction $\Psi_{0}\left( 
{\bf r}_{1},\cdots , {\bf r}_{\mbox{N}}\right)$ by
partial integration, or tracing out of the N-1 co-ordinates from the N-body
 pure density-matrix $\Psi^{\ast }_{0}\left({\bf r}_{1},\cdots , 
{\bf r}_{\mbox{N}}\right)\Psi_{0}\left({\bf r}^{\prime }_{1},\cdots , 
{\bf r}^{\prime }_{\mbox{N}}\right)$.  
(It is to be recalled in passing here that the condensate
 and the superfluid fractions are not the same thing. More
specifically, e.g., for the non-rotating ground state, 
the condensate fraction 
is generally less than unity while the superfluid fraction, that includes
 the condensate as well as the above-the-condensate fraction, is
exactly equal to unity. However, both the condensate as well as
 the superfluid fractions are characterized by the
 same quantum-mechanical phase whose gradient gives the superfluid 
 velocity. 
 Also, both vanish together above the critical temperature T$_{c}$. 
 In fact, it is the condensate that amplifies the effect of even the 
  weakest repulsive interaction leading to absence of single-particle
 excitation and giving rise to the phase rigidity).\\
\indent 
In this work, we have studied the effect of the 2-body 
 repulsive interaction
 on the condensate 
 fraction, and on the the 
critical angular velocities$\left(\Omega_{ci},\ i=0,1,2,\cdots \right)$ 
for the appearance of different vortex states for a quasi-2D Bose  
 gas confined in a highly anisotropic harmonic trap 
 $\left(\mbox{with }\omega_{z}\gg \omega_{\perp }\right)$ which is 
subjected to an overall rotation $\Omega $ about the z-axis. 
To this end we use
a variational approach and calculate the ground state energy,
the ground state wavefunction and the associated one-particle reduced 
density-matrix for different values of the 
interaction, 
 in given subspaces of the total quantized
angular momentum L$_{z}$.
The many-body variational ground state  
wavefunction is obtained through the exact diagonalization of the
 n$\times $n many-body Hamiltonian matrix with, e.g.,   
 n$=240782$ for L$_{z}$=N=15, using Davidson's algorithm.
A distinctive feature of the present work is that in constructing 
the many-body variational ground state wavefunction for a given
L$_{z}$ ($\equiv $ the
 eigenvalue corresponding to the z-component
 of the total angular momentum operator $\hat{\mbox{L}}_{z}$), 
we have included, in the configuration interaction, the one-particle
 states with the single-particle angular momentum$\left(\equiv \hat{\bf l}_{z}
\right)$ quantum number $m$ of {\em either} sign, as also the higher
 ``Landau-Level(LL)'' states.  
 One of our main results is the variation of the critical
 angular velocities $\Omega_{{\bf c}1}$ with the interaction strength 
 $\Lambda_{2}\equiv \left(\mbox{N-1}\right)\sqrt{\frac{2}{\pi }}
\frac{a_{sc}}{a_{z}}$(the meanings of various symbols will be given in 
the next section), namely, that not only does $\Omega_{c1}$ decrease with 
 increasing $\Lambda_{2}$, it also stays 
  systematically higher than its non-varional value, e.g., that given 
 by the relation $\Omega_{c1}=\omega_{\perp }\left(1-
\frac{\Lambda_{2}}{4}\right)$ 
for the weakly interacting, dilute case\cite{rokh99,wg00,bmot99}.
 We have also observed that the condensate fraction for the 
single vortex state with L$_{z}$=N(as also for the higher vortex states)
 does not change significantly
even if the 2-body scattering length $a_{sc}$ is changed over
 $\sim 4$ orders of magnitude in the moderately to the weakly
 interacting regime, namely, from $a_{sc}=1000 a_{0}$  
  to $1 a_{0}$, where $a_{0}$ is the Bohr radius.\\
\indent
This paper is organized as follows. In Section 2, we begin with a more 
general system to bring out, in passing, the mathematical analogy between
the system under study here and the other well known systems 
like in the Landau-Darwin-Fock problem  
 and argue that in a certain limiting case of interest to us
 here,    
it becomes essential to go beyond the Lowest Landau Level(LLL)
approximation so as to include higher LLs and with the single-particle
 angular momentum eigenvalue $m$ taking both positive
 as well as negative values in constructing the many-body 
 basis functions. Section 3 describes briefly the constrction of
the many-body basis functions and the determination of the
 variational ground state wavefunction by diagonalizing the
many-body Hamiltonian matrix using Davidson's algorithm. 
In Section 4, we outline the procedure
 for determining the critical angular velocities for the entry of
the vortices into the system, and the determination of various
 density profiles obtained from the one-particle reduced density-matrix
 characterizing the vortical state. Finally, in Section 5, we present
 and discuss our results, and end the paper with a brief conclusion.
\section{THE SYSTEM AND THE HAMILTONIAN}
\indent
We begin by considering the general case of a system of 
interacting, spinless, charged particles (bosons) confined in an external  
harmonic potential(trap). The 2-body interaction potential is, however,    
assumed to be gaussian in the particle-particle separation. The 
 trap is also subjected to an externally impressed rotation at an
 angular velocity 
${\bf \Omega }\equiv\Omega\hat{z}$, and to a uniform magnetic 
field ${\bf B}\equiv \mbox{B}\hat{z}$.
 The Hamiltonian for the system in a frame co-rotating with the   
 angular velocity ${\bf\Omega }$ is then
\begin{eqnarray}
\hat{\bf H}^{rot}&=&\hat{\bf H}^{lab}- 
{\bf \Omega }\cdot \hat{\bf L}^{lab} \nonumber \\
\hat{\bf H}^{lab}&=&
\sum_{i=1}^{\mbox{N}}\left[\frac{1}{2m} 
\left(\frac{\hbar }{i}{\bf\nabla }_{i} -q{\bf A}\left({\bf r}_{i}
\right)\right)^{2}
~+~{1\over 2}m\omega_{\perp }^{2}\left(r_{\perp i}^{2}+ 
 \lambda_{z}^{2} z_{i}^{2}\right)\right]
~+~ \nonumber \\ && 
{1\over 2}\cdot\frac{4\pi\hbar^{2}a_{sc}}{m}\cdot
\left({{1} \over {\sqrt{2\pi }\sigma}}\right)^{3}\sum_{i\neq j}
\exp^{-\frac{1}{2\sigma^{2}}
\left\{\left(r_{\perp i}-r_{\perp j}\right)^{2}~+~
       \left(z_{i}-z_{j}\right)^{2}\right\}}  \\
\mbox{with      }\ \ \   
\hat{\mbox{\bf L}}_{z}^{lab}&=& \sum_{i=1}^{\mbox{N}}
\hat{\mbox{\bf l}}^{lab}_{z\ i}
~=~\frac{\hbar }{i}\sum_{i=1}^{\mbox{N}}\  
\left({\bf r}_{i}\times {\bf \nabla }_{i} \right)_{z},\ \ \  
 {\bf B}=\nabla \times  {\bf A}\ \ \ \mbox{and }\ \ \   
 {\bf A}\equiv \frac{\mbox{B}}{2}\left({\hat{e}_{z}}\times {\bf r}\right) .
 \nonumber
\end{eqnarray}
 Here q is the charge of the particles and 
{\bf A}$\left({\bf r}\right)\equiv 
\frac{\mbox{B}}{2}\left(x\hat{e}_{y}-y\hat{e}_{x}\right)$  
is the electromagnetic vector potential in the symmetric gauge.
 The co-ordinates $\left\{ {\bf r}_{i} \right\}$ 
 and the corresponding canonical momenta 
$\left\{ \frac{\hbar }{i}{\bf \nabla }_{i} \right\}$ refer to the
laboratory frame. From now onwards, we will drop the superscript
 $lab$ on the Hamiltonian $\hat{\bf H}^{lab}$, the total angular momentum
 $\hat{\mbox{L}}_{z}^{lab}$, the single-particle angular momentum
 $\hat{\mbox{l}}_{z}^{lab}$ and their respective eigenvalues, and these will
 always be assumed to refer to the laboratory frame unless
 otherwise specified.\\
\indent   
We now make the following assumptions. The confining asymmetric 
 harmonic potential
 is highly oblate spheroidal, with 
 $\lambda_{z}\equiv\frac{\omega_{z}}{\omega_{\perp}}\gg 1$, and hence 
 our confined system is effectively quasi-2D with the x-y 
 rotational symmetry; the repulsive 2-body scattering is dominantly in the 
 s-wave channel with a scattering length $a_{sc}$ and   
 $\frac{4\pi\hbar^{2}a_{sc}}{m}$ having the dimension of Energy$\times$Volume.
 The range $\sigma $ of the 2-body interaction is small enough 
 compared to the inter-atomic spacing so as to effectively give a 
 $\delta $-function interaction potential
 $V\left({\bf r},{\bf r}^{\prime }\right)=\frac{4\pi\hbar^{2}a_{sc}}{m}\delta
\left({\bf r}-{\bf r}^{\prime }\right)$.\\ 
\indent
Now, for our N-body variational calculation we need to construct 
 the N-body basis-functions with proper symmetry. Since the system above
 is rotationally invariant in the x-y plane, the z-component of the
 total angular momentum(L$_{z}$) is a good quantum number leading to 
 block diagonalization of the Hamiltonian matrix into the subspaces of
 $\hat{\mbox{\bf L}}_{z}$. The N-body 
 basis-functions are, in turn, constructed 
 as linear combinations of the symmetrized products of a finite number 
 of single-particle basis-functions, 
 which are chosen to be eigenfunctions of the unperturbed 
single-particle Hamiltonian. 
 With this in mind, let us consider the non-interacting 
 single-particle Hamiltonian $\hat{\bf h}$    
 in the rotating frame, which we now separate into the z and the 
 (x-y)$\_$plane (commuting) components, $\hat{\bf h}_{z}$ and   
 $\hat{\bf h}_{\perp }$ respectively:
\begin{eqnarray}
\hat{\bf h}&=& \underbrace{\frac{1}{2m}\left(\frac{\hbar }{i}
{\bf\nabla }_{\perp }
 -q\frac{1}{2}\left(\mbox{B}\times {\bf r}_{\perp }\right)\right)^{2}
+{1\over 2}m\omega_{\perp }^{2}r_{\perp }^{2}-\frac{\hbar}{i}{\bf\Omega } 
\cdot\left({\bf r}_{\perp }\times {\bf \nabla }_{\perp }
\right)}_{{\bf h}_{\perp }}\nonumber \\ &+&
\underbrace{\frac{1}{2m}\left(\frac{\hbar }{i}{\bf\nabla }_{z}\right)^{2}
+ {1\over 2}m\omega_{z}^{2} z^{2}}_{{\bf h}_{z}}~\equiv~{\bf h}_{\perp }
 +{\bf h}_{z}.
\end{eqnarray}
The eigensolutions for $\hat{\bf h}_{z}$ are then given by 
\begin{eqnarray*}
\hat{\bf h}_{z} {\bf u}_{n_{z}}\left(z\right)&=&\epsilon_{n_{z}}
{\bf u}_{n_{z}}\left(z\right),\ \ \ \ \ \mbox{where}\ \ \ \ \ 
\epsilon_{n_{z}}=\left(n_{z}+{1\over 2}\right)\hbar\omega_{z},
 \ \ \ n_{z}=0,1,2, \cdots ,\\  
{\bf u}_{n_{z}}\left(z\right)&=&
\sqrt{{\alpha_{z}}\over {\sqrt{\pi }2^{n_{z}}
n_{z}!}}\ e^{-{1\over 2}\alpha_{z}^{2}z^{2}} \mbox{H}_{n_{z}}\left(
\alpha_{z}z\right), \\ 
\mbox{and }\ \ \ \ \  
\alpha_{z}&=&\sqrt{\frac{m\omega_{z}}{\hbar }}
\equiv \mbox{inverse of the longitudinal oscillator length}
\left(\mbox{a}_{z}\right)
\end{eqnarray*}
 Here H$_{n_{z}}$ is the Hermite polynomial. 
  We assume the system to be quasi-2D in that there is 
 practically no excitation along the relatively stiffer z-axis, and   
 hence we set $n_{z}=0$.\\
\indent
 Let us now consider $\hat{\bf h}_{\perp }$. Using 
 ${\bf \Omega } =\Omega \hat{e}_{z}$, $\mbox{B}=\mbox{B}\hat{e}_{z}$ and
 {\bf r}$_{\perp }=x\hat{e}_{x}+y\hat{e}_{y}$, we re-write 
 it more explicitly as
\begin{eqnarray}
\hat{\bf h}_{\perp }
&\equiv&\underbrace{-{{\hbar^{2}}\over {2m}}\left[{1\over r_{\perp }}
{{\partial }\over {\partial r_{\perp }}}\left(r_{\perp} {{\partial }\over 
{\partial r_{\perp}}} \right) +{1\over {r_{\perp }^{2}}}
{{\partial^{2} }\over {\partial \phi^{2}}}\right]+ 
{1\over 2}m\zeta_{\perp }^{2}r_{\perp }^{2}}_{\hat{\mathcal K}_{\perp } }
~-~\varsigma~\hat{\mbox{l}}_{z}~\equiv~\hat{\mathcal K}_{\perp }
~-~\varsigma~\hat{\mbox{l}}_{z} \\ 
\mbox{ with }\ \ {\bf \zeta }_{\perp }&\equiv & 
 \sqrt{ \left( \frac{\mbox{qB}}{2m}\right)^{2}
+\omega_{\perp }^{2}},\ \ \mbox{the frequency for the harmonic confinement 
in the x-y$\_$plane},
 \nonumber \\ \mbox{ and }\ \
{\bf \varsigma } &\equiv &\left(\frac{\mbox{qB}}{2m}+\Omega \right),\ \
\mbox{the cyclotron-rotational(or the centrifugal) angular velocity}
   \nonumber \\ 
 &&\ \ \ \ \ \ \ \ \ \ \ \ \ \ \ \ \ \ \mbox{about the z-axis}. 
 \nonumber
\end{eqnarray}
Here $\hat{\mathcal K}_{\perp }$ is the single-particle 2D-harmonic 
oscillator Hamiltonian, for which the eigensolutions are known to be: 
\begin{eqnarray}
\hat{\mathcal K}_{\perp }\mbox{u}_{n,m}\left(r_{\perp },\phi\right)&=&
\epsilon_{n,m}\mbox{u}_{n,m}\left(r_{\perp },\phi\right),\ \ \ \ \ \ \ \ \
\hat{\mbox{l}}_{z}\mbox{u}_{n,m}\left(r_{\perp },\phi\right)~=~
m\hbar \ \mbox{u}_{n,m}\left(r_{\perp },\phi\right),\nonumber \\
\mbox{with  }\ \ 
\mbox{u}_{n,m}\left(r_{\perp },\phi\right)&=&
\sqrt{\frac{\alpha_{\perp }^{2}}{\pi }\cdot\frac{\left({1\over 2}\left[
n-\mid m\mid \right] \right)!}{\left({1\over 2}\left[
n+\mid m\mid \right]\right)!}}\
\left(r_{\perp }\alpha_{\perp }\right)^{\mid m\mid }\
 e^{-{1\over 2}\alpha_{\perp }^{2} r_{\perp }^{2}}\
e^{im\phi }\ \mbox{L}^{\mid m\mid }_{{1\over 2}\left(n-\mid m\mid \right)}
\left(\alpha_{\perp }^{2}r_{\perp }^{2}\right), \nonumber \\ 
\mbox{where} \ \ \ \ \ \alpha_{\perp }&=&  
\sqrt{\frac{m\zeta_{\perp }}{\hbar }}\equiv \mbox{the inverse of the
 transverse oscillator length}\left(\mbox{a}_{\perp }\right), \nonumber \\
\mbox{and}\ \ \ \ \   
\epsilon_{n,m}&=&\left(\underbrace{2n_{r}+\mid m\mid}_{n}+1\right)
\hbar\zeta_{\perp }
\equiv\left(n+1\right)\hbar\zeta_{\perp }, \nonumber \\
\mbox{with}\ \ \ &&
n_{r}=0,1,2,\cdots ,\ \ 
m=-\infty ,\cdots , -1,0,+1,\cdots , +\infty , \nonumber \\
\mbox{ or equivalently,  } &&
n=0,\ 1,\ 2,\ \cdots ,\ \  
m = +n,\ +n-2,\cdots ,-n+2,\ -n.
\end{eqnarray}
Here  
L$^{\mid m\mid }_{{1\over 2}\left(n-\mid m\mid \right)}
\left(\alpha_{\perp }^{2}r_{\perp}^{2}\right)$ 
 is the Associated Laguerre polynomial. It may be noted in passing
 that the above Hamiltonian represents the well known 
 Landau-Darwin-Fock \cite{foc28,dar30}
 problem in which $\zeta_{\perp }$ is the frequency for the  
 harmonic confining potential in the x-y$\_$plane and 
 ${\bf \varsigma }$ is the cyclotron-rotational
 (centrifugal) angular velocity about the z-axis.   
Limiting to $n_{r}=0$ and taking $m=0,+1,+2,+3,\cdots ,$ corresponds to   
 the ``Lowest-Landau Level(LLL)'' approximation.\\
\indent
 Let us examine the centrifugal/mechanical 
 stability of the above system by re-writing $\hat{\bf h}_{\perp }$ as 
\begin{eqnarray*}
\hat{\bf h}_{\perp }&=&\underbrace{
\frac{1}{2m}\left(\frac{\hbar }{i}{\bf\nabla }_{\perp }
 -q\frac{1}{2}\left(\mbox{B}\times{\bf r}_{\perp }\right)
 -m\left(\Omega\times {\bf r}_{\perp }\right)\right)^{2}}_{\hat{\cal T}}
  \nonumber \\ 
 &+&\underbrace{{1\over 2}m\omega_{\perp }^{2}r_{\perp }^{2}-
\frac{1}{2}m \left(\Omega \times {\bf r}_{\perp }\right)^{2}
-q\frac{1}{2}\left(\mbox{B}\times{\bf r}_{\perp }\right)
\cdot\left(\Omega\times {\bf r}_{\perp }\right)}_{\hat{\cal U}}
 ~\equiv~\hat{\cal T}+\hat{\cal U} 
\end{eqnarray*}
Here $\hat{\mathcal T}$ is clearly a positive-definite operator. 
 It can readily
 be shown that $\hat{\cal U}$ is negative-definite, null or  
 positive definite according as
$\left(\zeta_{\perp }+\varsigma\right)
 \left(\zeta_{\perp }-\varsigma\right)$     
 is negative , zero or positive, respectively.\\  
\indent
For $\hat{\cal U}$ negative-definite,
 the Hamiltonian 
 $\hat{\bf h}_{\perp }\equiv \hat{\cal T}+\hat{\cal U}$ is unbounded 
 from below and there are no stable solutions.
 In the simpler case of vanishing magnetic filed B=0, this
 situation arises for $\omega_{\perp }< \Omega $, i.e, when
 the rotational angular velocity $\Omega $ becomes larger than the 
 confining harmonic trap frequency $\omega_{\perp }$.\\
 \indent 
For the special case of $\hat{\mathcal U}$ null, we have 
$\zeta_{\perp }=\varsigma $ and the Hamiltonian reduces to   
$\hat{\bf h}_{\perp }={\mathcal K}_{\perp }
~-~\zeta_{\perp }~\hat{\mbox{l}}_{z} \ .$
This gives rise to a situation analogous to the Landau problem 
when the frequency for the harmonic confinement is equal
to the centrifugal frequency.
 Setting $\zeta_{\perp }=\frac{1}{2}\omega_{c}$, 
 we have the eigenvalue equation
\begin{eqnarray}
&&\hat{\bf h}_{\perp }{\bf u}_{n,m}\left(r_{\perp },\phi\right)=
\left(\hat{\mathcal K}_{\perp }-\zeta_{\perp }{\hat{\mbox{l}}_{z}}\right)
{\bf u}_{n,m}\left(r_{\perp },\phi\right)=
\left(\epsilon_{n,m}-m\hbar\zeta_{\perp }\right)
{\bf u}_{n,m}\left(r_{\perp },\phi\right) \nonumber \\ 
&\mbox{with }&\left(\epsilon_{n,m}-m\hbar\zeta_{\perp }\right)=
\left(\underbrace{n_{r}+\frac{1}{2}\left(\mid m\mid-m\right)}_{\cal N}
+\frac{1}{2}\right)\hbar\omega_{c}\equiv 
\left({\cal N}+\frac{1}{2}\right)\hbar\omega_{c}\nonumber \\  
&\mbox{where}&\ n_{r}=0,1,2,3,\cdots ,\ \mbox{and}  
\ \ m=-\infty \cdots -2,-1,0,+1,+2,\cdots +\infty ; \nonumber \\
 &\mbox{ or }&\mbox{equivalently }\ {\cal N}=0,1,2,3,\cdots , \nonumber \\
 &&\ \mbox{and} \ \ \ \ \  
 \ m=-{\cal N},-\left({\cal N}-1\right), \cdots ,-2,-1,0,+1,+2,
 \cdots , +\infty . 
\label{llseq}
\end{eqnarray}
Each of the ${\cal N}$ levels, the so-called Landau Levels(LL), is  
infinitely degenerate corresponding to the infinitely many possible 
 values of $m$. 
It is clear from the ordering of the single-particle 
energy levels (without interaction) in equation~(\ref{llseq}) that 
 the single-particle
 states with positive $m$ values (i.e those with the angular momentum 
 parallel to the overall rotational angular velocity ${\mathbf \varsigma}
 \equiv \varsigma \hat{e}_{z}$) are 
 energetically favored. These states constitute a massively degenerate 
 manifold. This is the usual rationale for using the
 positive $m$ values only, in constructing the variational wavefunction
 for the Landau-like problem. 
This degeneracy for the special case of $\zeta_{\perp }=\varsigma $ is, 
however, lifted by the inter-particle interactions.\\
\indent 
Finally, we consider the physically interesting case of 
$\hat{\mathcal U}$ positive-definite. 
 The single-particle non-interacting Hamiltonian $\hat{\bf h}_{\perp }$,
 now becomes  
$\hat{\bf h}_{\perp }=
{\mathcal K}_{\perp } ~-~\varsigma~\hat{\mbox{l}}_{z}$
with the eigenvalue solution:
\begin{eqnarray}
&&\hat{\bf h}_{\perp }{\bf u}_{n,m}\left(r_{\perp },\phi\right)=
\left(\hat{\mathcal K}_{\perp }-\varsigma 
\hat{\mbox{l}}_{z}\right){\bf u}_{n,m}\left(r_{\perp },\phi\right)=
\left(\epsilon_{n,m}-m\hbar \varsigma\right)
{\bf u}_{n,m}\left(r_{\perp },\phi\right)\nonumber \\ 
&&\left(\epsilon_{n,m}-m\hbar \varsigma\right)
=2\left(n_{r}+\frac{1}{2}\left(\mid m\mid -
m\frac{\varsigma}{\zeta_{\perp }}\right)
+\frac{1}{2}\right) \hbar\zeta_{\perp }, \nonumber \\
\mbox{where }&& n_{r}=0,1,2,\cdots ,\ \ \ \mbox{and }\ \ \ 
m=-\infty, \cdots -2,-1,0,+1,+2,\cdots +\infty .
\label{e2p7}
\end{eqnarray}
In this physically relevant situation, the particle as observed
 in the rotating frame finds itself in a 
shallower harmonic potential 
of frequency $\sqrt{\zeta_{\perp }^{2}-\varsigma^{2}}$,  
and the states will be more spread out. It can be seen from
 equation(\ref{e2p7}) that 
 for the centrifugal frequency $\varsigma $  
 significantly smaller than the confining frequency $\zeta_{\perp }$,   
 the degeneracy of the Landau levels is lifted even without interaction. 
 Further, the interaction between
 the particles causes the the different single-particle angular momentum 
 states to scatter into each other. 
Thus, for the slowly rotating systems
 and for moderately, and a fortiori for strongly interacting systems, 
 it may be  
 necessary to include single-particle basis functions with different 
 values of $n_{r}$ and with  
 the angular momentum quantum number $m$ 
 taking both positive and negative values. This is the case
 we shall be concerned with in the following with B=0 and $q=0$. 
\section{THE MANY-BODY VARIATIONAL WAVEFUNCTION}
\indent
We need to perform the diagonalization of the Hamiltonian matrix 
in the subspaces of $\hat{\mbox{\bf L}}_{z}$ only. 
The N-body variational wavefunction $\Psi\left({\bf r}_{1},\cdots ,
{\bf r}_{\mbox{N}}\right)$ 
 is a linear combination of the symmetrized products
$\psi_{b} \left({\bf r}_{1},\ {\bf r}_{2},
 \cdots , {\bf r}_{\mbox{N}} \right) $ of the one-particle
basis functions $u_{\underbrace{n,m,n_{z}}_{\bf n}}
\left({\bf r}\right) \equiv u_{\bf n}\left({\bf r}\right)$
  introduced earlier:
\begin{eqnarray*}
\Psi\left({\bf r}_{1},\ {\bf r}_{2}, \cdots ,{\bf r}_{\mbox{N}}
\right)&=&\sum_{b}\ C_{b}\  \psi_{b} \left({\bf r}_{1},\ {\bf r}_{2}, 
 \cdots , {\bf r}_{\mbox{N}} \right)  \\
\mbox{with }\ \ \ \psi_{b} \left({\bf r}_{1},\ {\bf r}_{2},
 \cdots , {\bf r}_{\mbox{N}} \right)&=&\frac{1}{\sqrt{\mbox{N!}}}
\sum_{\mbox{P}}\mbox{P}
\left(\frac{1}{\sqrt{\nu_{\bf 0}!}}\prod_{i=1}^{\nu_{\bf 0}}
u_{\bf 0}\left({\bf r}_{i}\right)\cdot
\frac{1}{\sqrt{\nu_{\bf 1}!}}\ \prod_{i=\nu_{\bf 0}+1}^{
\nu_{\bf 0}+\nu_{\bf 1}} u_{\bf 1}\left({\bf r}_{i}\right)\cdots \right. \\
&&\hspace{100pt}\left.\cdots\frac{1}{\sqrt{\nu_{\bf k}!}}\ 
\prod_{i=\nu_{\bf 0}+\nu_{\bf 1}+\cdots +\nu_{{\bf k}-1}+1}^{
\nu_{\bf 0}+\nu_{\bf 1}+\cdots +\nu_{{\bf k}-1}+\nu_{\bf k}}
u_{\bf k}\left({\bf r}_{i}\right) \right),
\end{eqnarray*}
\begin{eqnarray*}
b&\equiv &\left(\nu_{\bf 0},\ \nu_{\bf 1},\ \cdots, \nu_{\bf j},\ \cdots,
 \ \nu_{\bf k}\right), \ \ \ \ \ \ \
\sum_{{\bf j}={\bf 0}}^{\bf k}\nu_{\bf j}=\mbox{N},\ \ \ \ \ \ 
\sum_{{\bf j}={\bf 0}}^{\bf k}\nu_{\bf j}m_{\bf j}
=\mbox{L}_{z}.
\end{eqnarray*}
Here $\left\{C_{b} \right\}$ are the variational parameters, and 
 P permutes the N particle co-ordinates. 
 Also, $\nu_{\bf j}$ is the occupancy of the {\bf j}th 
single-particle basis function$\left(u_{\bf j}\right)$.  
The many-body quantum index $b$, labelling the many-body basis
function $\psi_{b} \left({\bf r}_{1},
 \cdots , {\bf r}_{\mbox{N}} \right)$, stands for a set of 
 single-particle quantum numbers required to satisfy the above
 two defining relations between the single-particle occupation quantum 
 numbers $\left\{\nu_{\bf j}\right\}$, the single-particle
 angular momentum quantum numbers
 $\left\{m_{\bf j}\right\}$, the number of particles N 
 and the total angular momentum L$_{z}$. 
 At this point it becomes more convenient to switch over to 
 second-quantization notation in the occupation number
$\left(\nu_{\bf j}\right)$ representation. The basis function is then  
\begin{eqnarray}
\left|\left.\psi_{b}\right\rangle\right. &\equiv &
{{\left(a^{\dagger }_{\bf 0}\right)^{\nu_{\bf 0}} 
\left(a^{\dagger }_{\bf 1}\right)^{\nu_{\bf 1}}
\left(a^{\dagger }_{\bf 2}\right)^{\nu_{\bf 2}}\cdots 
\left(a^{\dagger }_{\bf k}
\right)^{\nu_{\bf k}} } \over 
{\sqrt{\nu_{\bf 0}!\ \nu_{\bf 1}!\ \nu_{\bf 2}!\cdots \nu_{\bf k}}
 }}\left| \left. 0\ \right\rangle\right. \hskip 1cm \mbox{with} \hskip 1cm
\sum_{{\bf j}={\bf 0}}^{\bf k}\nu_{\bf j}={\mbox{N}}, \ \ 
\sum_{{\bf j}={\bf 0}}^{\bf k}\nu_{\bf j}m_{\bf j}={\mbox{L}_{z}} \nonumber\\
&\equiv & 
\left|\left. \left(\nu_{\bf 0}\ \nu_{\bf 1}\ \cdots \ \nu_{\bf k}\right) 
:\ \sum_{{\bf j}={\bf 0}}^{\bf k}\nu_{\bf j}={\mbox{N}},\
\sum_{{\bf j}={\bf 0}}^{\bf k}\nu_{\bf j}m_{\bf j}={\mbox{L}_{z}}
 \right\rangle\right. , 
\label{2p8}
\end{eqnarray}
and the many-body Hamiltonian
 written in the second-quantized notation is
\begin{eqnarray*}
{\bf H}&=&
 \sum_{{\bf i} ,\ {\bf j}}\left\langle {\bf i}\left|~1~
\right| {\bf j}\right\rangle
a_{\bf i}^{\dagger }a_{{\bf j}} +{1\over 2}
\sum_{{\bf i}_{1},\ {\bf j}_{1}}\ \sum_{{\bf i}_{2},\ {\bf j}_{2}}
\left\langle {\bf i}_{1}, {\bf i}_{2} \left|~2~\right|
 {\bf j}_{1}, {\bf j}_{2} \right\rangle 
\left(a_{{\bf i}_1 }^{\dagger }a_{{\bf j}_1 }a_{{\bf i}_2 }^{\dagger }
a_{{\bf j}_2 }
-\delta_{{\bf i}_{2}{\bf j}_{1}}a_{{\bf i}_1 }^{\dagger }a_{{\bf j}_2 }\right)
\end{eqnarray*}
Here $a^{\dagger }_{\bf j}\left(a_{\bf j}\right)$ are the usual bosonic
 creation(annihilation) operators, and  
$\left\langle {\bf i}\left|~1~\right|{\bf j}\right\rangle $ and 
$\left\langle {\bf i}_{1}, {\bf i}_{2}\left|~2~\right|{\bf j}_{1}, {\bf j}_{2}
\right\rangle$ are the one-body and the two-body matrix-elements 
 respectively over the single-particle basis functions.
 The evaluation of matrix-elements for the kinetic energy and
 the harmonic trapping potential over the single-particle basis chosen 
 in the previous section is trivial. It is also possible to obtain  
 closed form expressions for the matrix-elements for the
two-body (as also for the n-body, n=2,3,$\cdots$) gaussian potential(s) over 
the above basis as they
reduce to multi-dimensional gaussian integrals. 
 (In the calculations presented here, however, 
 we have not considered the three or the higher-body potentials).\\
\indent
 The variational parameters $\left\{C^{0}_{b}\right\}$ for the
 ground state wavefunction $\Psi_{0}=\sum_{b}C^{0}_{b}\psi_{b}$ are 
now determined by minimizing
$\mbox{K}_{0}\left\{\Psi_{0}\right\}\equiv
\left\langle \Psi_{0} \left|{\bf H}\right| \Psi_{0}\right\rangle 
 -\mbox{E}_{0} \left\langle \Psi_{0}\left.\right| \Psi_{0}\right\rangle $  
with respect to $\Psi_{0}$. The Lagrange multiplier E$_{0}$ is identified
 as the variational energy for the ground state. 
The $i$th excited state $\Psi_{i}=
\sum_{b}C^{i}_{b}\psi_{b}$ is determined by
 carrying out the variational minimization in the restricted Hilbert
 subspace which is orthogonal to the $(i-1)$ states determined 
earlier, i.e. by minimizing   
$\mbox{K}_{i}\left\{\Psi_{i}\right\}\equiv
\left\langle \Psi_{i} \left|{\bf H}\right| \Psi_{i}\right\rangle 
 -\mbox{E}_{i} \left\langle \Psi_{i}\left.\right| \Psi_{i}\right\rangle
\ \ \ \ \mbox{with } \ \ 
\left\langle \Psi_{j}\left.\right| \Psi_{i}\right\rangle =0
\ \ \mbox{for }\ \ j=0,1,\cdots , (i-1).$\\
\indent
The Davidson algorithm of iterative diagonalization \cite{dav75} 
 is based on a procedure where one keeps a minimum set of $i$  
 orthogonal, trial wavefunctions 
which spans a small subspace ${\cal S}_{i}$ of the full many-body Hilbert
 space: 
\begin{eqnarray*}
{\cal S}_{i}\equiv \left\{\Psi_{j}\left|\right.
\Psi_{j}=\sum_{b}C^{j}_{b}\psi_{b}\ ,\ \ j=0,1,\cdots i\ \ \mbox{and} 
\left\langle \Psi_{j}\left.\right| \Psi_{k}\right\rangle =\delta_{jk}\ \  
\mbox{for }\ j,k=0,1,\cdots ,i \right\}. 
\end{eqnarray*}
This small subspace ${\cal S}_{i}$ is chosen to contain  
 dominant contributions from the ground state and the first few 
excited states. An $i\times i$ representation $H^{\left(i\right)}$ of the 
 Hamiltonian
 {\bf H} is obtained over this small subspace to set up the 
  small eigenvalue equation $H^{\left(i\right)}{\bf a}_{\nu }=
\lambda^{i}_{\nu }{\bf a}_{\nu },\  \nu =0,1,2,\cdots i$.  
  $H^{\left(i\right)}$ is an  
  effective Hamiltonian for the system over the small subspace 
 ${\cal S}_{i}$
 spanned by $\left\{\Psi_{j},\  j=0,1,2,\cdots , i\right\}$. 
The eigenvalue $\lambda_{\nu }^{i} $ is the $\nu $th approximate 
eigenvalue. The
convergence for the $\nu $th state is achieved when the residual vector 
 for the $\nu$th state $\left|\left. \Delta\Psi_{\nu }\right\rangle\right.=
{\bf H}\left|\left.\Psi_{\nu }\right\rangle\right. - \lambda_{\nu }^{i}
 \left|\left.\Psi_{\nu }\right\rangle\right.$ becomes a null vector.
 If the convergence has not been achieved, the residual
 vector $\Delta\Psi_{\nu }$, after orthonormalization,
 is added to the list of trial vectors 
 to augment the subspace ${\cal S}_{i}$
 to obtain ${\cal S}_{i+1}=
\left\{\Psi_{j},\  j=0,1,2,\cdots , i+1\right\}$. The procedure
is continued till the convergence is obtained. In the process, if the 
size of the 
subspace ${\cal S}_{i}$ becomes unwieldingly large, a certain number of higher
 eigenvectors are dropped from ${\cal S}_{i}$ and the procedure
 once again initiated.\\
\indent   
For N=15 particles, for example, we have carried out calculations 
 for all the total angular momentum states in the regime 
 $0\leq \mbox{L}_{z}\leq 3\mbox{N}$. 
Diagonalization of the Hamiltonian matrix is performed for
each of the subspaces of L$_{z}$ separately. (We have set $n_{z}=0$ in
 the single-particle basis function $u_{n_{z}}\left(z\right)$ 
 as discussed earlier.)
 The single-particle basis
$ u_{n,m}\left(r_{\perp },\phi\right)$ with $n\equiv 2n_{r}+\mid m\mid$
 and $n_{r}=0,1,\cdots ,\ \& \mid m\mid=0,1,2,\cdots ,$ 
 spanning the 2D x-y plane for a given subspace of 
 L$_{z}$, is chosen  
 as follows. It is convenient to define  
 $l\equiv\left[\frac{\mbox{L}_{z}}{\mbox{N}}\right]$, where for real
 $x$, the symbol $\left[x\right]$ denotes the greatest integer less than
 or equal to $x$. For very weakly interacting 
 particles, almost all the particles will condense into a single
 one-particle state, $n=m=l$, a Yrast-like state. 
 As the interaction becomes stronger, the particles
 start scattering out to other single-particle states around this 
 state, {\em i.e.} 
 some of the particles go to states with higher angular momentum 
 while some of them go to states with lower angular 
 momentum(the two-body interaction conserves the total angular momentum). 
 The single-particle angular momentum for the basis functions is now 
 chosen to be: $m=l-n_{b},\ l-n_{b}+1,\ \cdots l+n_{b}-1,\ l+n_{b}$,  
 where $n_{b}$ is some positive integer that we have chosen to be 3, 4  
 or more depending on the strength of the interaction and the computational 
 resources available($n_{b}$ is a kind of the size of the single-particle
 basis chosen for the calculation for a given value of L$_{z}$ and describes
 configurational interaction). In all our  
 calculations presented here we have taken $n_{r}=0,\  1,$ and
 $n_{b}=3$. Thus for example, for N=15 and for the chosen   
  subspace L$_{z}=33$, we get $l\equiv\left[\frac{\mbox{L}_{z}}{\mbox{N}}
 \right]=2$, and the 
 single-particle angular momentum quantum number takes values $m=-1,0,+1,+2,
 +3,+4,+5$. Then, with $n_{r}=0,1$, the single-particle basis set turns
 out to be 
\begin{eqnarray*}
\left\{ u_{0,0},\ u_{1,+1},\ u_{2,+2},\ u_{3,+3},\ u_{4,+4},\ u_{5,+5},\
u_{1,-1},\ u_{2,0},\ u_{3,+1},\ u_{4,+2},\ u_{5,+3},\ u_{3,-1} 
\right\}. 
\end{eqnarray*}
Thus, the N(=15)-body basis functions $\left\{\psi_{b}\right\}$, for  
 the L$_{z}=33$ subspace, are to be constructed from these
 single-particle basis functions. These were used in turn to 
 construct the variational trial function $\Psi =\sum_{b}C_{b}\psi_{b}$, 
 and hence the ground state properties for the given value of L$_{z}$.     
\section{CRITICAL ANGULAR VELOCITIES AND DENSITY PROFILES FOR THE
CONFINED ROTATING BOSE GAS AT ZERO TEMPERATURE}
\indent
From the variational ground state wavefunction $\Psi_{0}\left({\bf r}_{1},
\cdots , {\bf r}_{\mbox{N}} \right)$ obtained in the previous section,
 we now, in the following, go on to calculate various quantities of interest.\\
\indent
Let us first calculate the critical angular velocities  
 $\left\{ \Omega_{{\bf c}i}\right\}$ for the onset of different
vortical states. Note that for the  
 system of N particles confined in a trap rotating at an angular
 velocity $\Omega $, the thermodynamic
equilibrium corresponds to the minimum of the free energy F given by
$\exp\left\{-\beta \mbox{F}\right\}=\mbox{Tr}\left(\exp\left\{-
\beta\left(\hat{\bf\mbox{H}}^{lab}
-\hbar\Omega\hat{\mbox{L}}^{lab}_{z}\right)\right\} \right),$
where $\hat{\bf\mbox{H}}^{lab}-\hbar\Omega\hat{\bf\mbox{\bf L}}^{lab}_{z}
\equiv \hat{\bf\mbox{H}}^{rot}\left(\Omega \right)$,  
 as we have noted earlier, is the Hamiltonian of the system in the
 co-rotating frame. For a system at T=0, the expression 
 for the free energy reduces to a more simpler form
$\mbox{F}=\left\langle\Psi_{0}\left|\left(\hat{\bf \mbox{H}}^{lab}
-\hbar\Omega \hat{\mbox{\bf L}}^{lab}_{z}\right)\right|\Psi_{0}\right\rangle
=\left\langle\hat{\bf\mbox{H}}^{lab}\right\rangle
-\hbar\Omega\left\langle\hat{\mbox{\bf L}}_{z}^{lab}\right\rangle$
where $\Psi_{0}$ is the ground state wavefunction of the system 
 in the laboratory frame for a given value of L$_{z}$ obtained,  
 in our case, variationally.
The above relation can as well be seen as the minimization
of the energy $\left\langle{\bf\hat{\mbox{H}}}^{lab}\right\rangle
\equiv \mbox{E}^{lab}_{0}$ 
subject to the constraint that the system has the angular
momentum expectation value $\left\langle{\bf\hat{\mbox{L}}}_{z}^{lab}
\right\rangle$.  
 The angular velocity $\Omega $ is then the corresponding 
Lagrange multiplier. Since we have constructed our variational
wavefunction $\Psi_{0}$ to be an eigenfunction of the total angular 
 momentum operator
 ${\bf\hat{\mbox{L}}}_{z}^{lab}$,  we will necessarily have
 $\left\langle{\bf\hat{\mbox{L}}}_{z}^{lab}\right\rangle=\mbox{L}_{z}$. 
 Initially, when the system is subjected to
no external rotation, the angular momentum state 
L$_{z}=0$ corresponds to the ground state of the
system. As the system is rotated, other non-zero angular
momentum states $\left(\mbox{L}_{z}\neq 0\right)$ successively  
become the ground state of the system. 
These are obtained by minimizing the energy 
$\mbox{E}^{rot}_{0}\left(\mbox{L}_{z},\Omega,\Lambda_{2}\right)$ 
in the rotating frame: 
\begin{eqnarray}
\mbox{E}^{rot}_{0}
\left(\mbox{L}_{z},\Omega,\Lambda_{2}\right)\equiv
\mbox{E}_{0}^{lab}\left(\mbox{L}_{z},\Lambda_{2}\right)-
 \hbar\Omega\mbox{L}^{lab }_{z} 
\label{4p3} 
\end{eqnarray}
Here $\Lambda_{2}$ measures the two-body interaction strength and has 
been defined in an earlier section.
Thus, the critical angular velocity $\Omega_{c}$ 
 beyond which the higher angular momentum state L$_{z}$, say, becomes  
 lower in energy in the rotating frame compared to the lower angular 
 momentum state L$_{z}^{\prime }
\left(<\mbox{L}_{z}\right)$ is given by: 
\begin{eqnarray}
\Omega_{c}
=\frac{\mbox{E}^{lab}_{0}\left(\mbox{L}_{\mbox{z}},
\ \Lambda_{2}\right)-\mbox{E}^{lab}_{0}
\left(\mbox{L}^{\prime }_{\mbox{z}},
\ \Lambda_{2}\right)}
{\left(\mbox{L}_{\mbox{z}}-
\mbox{L}^{\prime }_{\mbox{z}}\right)\hbar }
\label{4p4}
\end{eqnarray}
 This gives us a number of critical angular velocities 
$\left\{\Omega_{{\bf c}i},\ i=0,1,2,\cdots\right\}$ at which the
 ground state of the rotating system changes its angular momentum 
 state abruptly. The angular
momentum $\mbox{L}_{z}$ of the ground state {\em vs} 
the angular velocity $ \Omega $ relation,   
called the L$_{z}-\Omega $-phase diagram, or the stability line, for
  the rotating system was calculated in terms
 of the variational ground states obtained. We present this 
 in Fig. 1.\\     
\indent
Next, we calculate the density profile and the circulation  for 
the vortex states. For this we need the single-particle 
reduced density-matrix.   
From the many-body ground state wavefunction 
$\Psi_{0}\left({\bf r}_{1},
\cdots ,{\bf r}_{\mbox{N}}\right)$, obtained through the variational  
exact diagonalization(ED),  
the one-particle reduced density matrix
$\rho_{1}\left({\bf r},{\bf r}^{\prime }\right)$ is obtained
by integrating out the N-1 co-ordinates:
\begin{eqnarray}
\rho_{1}\left({\bf r},{\bf r}^{\prime }\right)&\equiv&\int\int\cdots\int
 d{\bf r}_{2} \ {\bf r}_{3}
\cdots {\bf r}_{\mbox{N}} \Psi_{0}^{\ast }\left({\bf r},{\bf r}_{2},
{\bf r}_{2},
\cdots ,{\bf r}_{\mbox{N}}\right)\Psi_{0}\left({\bf r}^{\prime },{\bf r}_{2},
{\bf r}_{3},\cdots ,{\bf r}_{\mbox{N}}\right) \nonumber \\
&=&\sum_{n}\sum_{m}\sum_{n_{z}}\sum_{n^{\prime }}\sum_{m^{\prime }}
\sum_{n^{\prime }_{z}}\rho_{nmn_{z},n^{\prime }
m^{\prime }n^{\prime }_{z}}\ u^{\ast }_{n,m,n_{z}}\left({\bf r}\right)
u_{n^{\prime },m^{\prime },n^{\prime }_{z}}
\left({\bf r}^{\prime }\right) \nonumber \\ 
&\equiv &\sum_{\mu }\lambda_{\mu }\ \chi^{\ast }_{\mu }\left({\bf r}\right)
\chi_{\mu }\left({\bf r}^{\prime }\right), \nonumber \\
 \mbox{with}\ \ \ \ \ \ \ \ \  
\chi_{\mu }\left({\bf r}\right)&\equiv &\sum_{n}\sum_{m}\sum_{n_{z}} 
c^{\mu }_{n,m,n_{z}}u_{n,m,n_{z}}\left({\bf r}\right)
\ \ \mbox{and}\ \ \ \ \sum_{\mu }\lambda_{\mu }=1,
\label{2p11}
\end{eqnarray}
where $\left\{\lambda_{\mu }\right\}$ are the eigenvalues and 
$\left\{\chi_{\mu }\left({\bf r}\right)\right\}$  
 the corresponding eigenvectors of the one-particle reduced 
density-matrix $\rho_{1}\left({\bf r},{\bf r}^{\prime }\right)$. 
The diagonal
part $\rho_{1}\left({\bf r},{\bf r}\right)\equiv \rho\left({\bf r}\right)$ 
gives the single-particle
density-profile of the system(the condensate $+$ the non-rotating 
 fraction $+$ the 
above-the-condensate fraction).
For our finite system, the BEC corresponds to a single 
eigenvalue $\lambda_{\mu}$ 
 being significantly larger than the rest of the eigenvalues.
As noted before, our system is quasi-2D, {\em i.e.}, there is 
practically no excitation in the z-direction. So we take n$_{z}=0$ 
and the summation over $n_{z}$ goes off. Also, we note that  
 $\chi_{\mu }\left({\bf r}\right)$ is an eigenvector of
 the single-particle angular momentum
  $\hat{\mbox{\bf l}}_{z}$ with eigenvalue $m_{\mu }$, 
  hence the summation over $m$ 
in equation(\ref{2p11}) too goes away. The only 
summation that we are left with is then over $n_{\mu }\left(\equiv 
2n_{r \mu}+\left| m_{\mu }\right| ,\ \mbox{where}\ n_{r \mu}=0,1,2,\cdots 
\right)$.  
Thus we get
\begin{eqnarray}
\chi_{\mu }\left({\bf r}\right)&=&\sum_{n_{\mu }=\mid m_{\mu }\mid,\  
\mid m_{\mu }\mid +2, \cdots }c^{\mu }_{n_{\mu },m_{\mu },0}\ 
u_{n_{\mu },m_{\mu },0} 
\left({\bf r}\right) 
\end{eqnarray}
Note that n$_{\mu }$ increases in steps of 2, which comes
from quantization.
 Tracing out the z co-ordinate, the density $\rho(r_{\perp },\phi)$, the
 current $j_{\phi }(r_{\perp },\phi)$, the velocity 
$v_{\phi }(r_{\perp },\phi)$ and the circulation $\kappa(r_{\perp })$
 profiles  
 in the x-y plane, 
 for the quasi-2D system, turn out to be  
\begin{eqnarray}
\rho\left(r_{\perp },\phi\right)&=&\frac{\alpha_{\perp }^{2}}{\pi }\cdot
 e^{-\alpha_{\perp }^{2}r_{\perp }^{2}}
 \sum_{\mu }\ \lambda_{\mu } \nonumber \\ && 
{\left| \sum_{n_{\mu }=\mid m_{\mu }\mid,\ \mid m_{\mu }\mid +2, \cdots }   
 \ c^{\mu }_{n_{\mu },m_{\mu },0}\ 
 \sqrt{\frac{\left({1\over 2}\left[n_{\mu }-\mid m_{\mu }\mid \right]\right)!}
{\left({1\over 2}\left[n_{\mu }+\mid m_{\mu }\mid \right]\right)!}} 
\left(r_{\perp }\alpha_{\perp }\right)^{\mid m_{\mu }\mid }
\mbox{L}^{\mid m_{\mu }\mid }_{{1\over 2}
\left(n_{\mu }-\mid m_{\mu }\mid \right)}  
\left(r_{\perp }^{2}\alpha_{\perp }^{2}\right)\right|}^{2}
 \nonumber \\  &\equiv & \rho\left(r_{\perp }\right),\\
\mbox{j}_{\phi }\left(r_{\perp},\phi\right)&=&
\sqrt{\frac{\hbar\omega_{\perp}}{m}}\cdot
\left(\frac{1}{\alpha_{\perp }r_{\perp }}\right) 
\cdot\sum_{\mu }\ m_{\mu }\
 \lambda_{\mu }\  {\left| \chi_{\mu }\left(r_{\perp},\phi\right)\right|}^{2}
 \ \ \ \equiv \ \mbox{j}_{\phi }\left(r_{\perp }\right),\ \\ 
\mbox{and }~&&\mbox{j}_{r_{\perp }}\left(r_{\perp},\phi\right)=
\mbox{j}_{z}\left(r_{\perp},\phi\right)=0;\nonumber \\
 \mbox{v}_{\phi }\left(r_{\perp},\phi\right)&=&
\sqrt{\frac{\hbar\omega_{\perp }}{m}}\cdot
\left(\frac{1}{\alpha_{\perp }r_{\perp }}\right)\cdot
\sum_{\mu }\ m_{\mu }\ \lambda_{\mu }\ 
\frac{\left| \chi_{\mu }\left(r_{\perp},\phi\right)\right|^{2}}
{\rho\left(r_{\perp},\phi\right)}\ \ 
 \equiv \mbox{v}_{\phi }\left(r_{\perp }\right),\\ \mbox{and}~&&
\mbox{v}_{r_{\perp }}\left(r_{\perp},\phi\right)=
\mbox{v}_{z}\left(r_{\perp},\phi\right) = 0;\nonumber \\
\kappa(r_{\perp})&\equiv& \oint_{r_{\perp }=constant} {\bf v}\cdot d{\bf l}=
\int_{0}^{2\pi }\mbox{v}_{\phi }\left(r_{\perp }\right)\ r_{\perp }d\phi
=2\pi \ r_{\perp }\ \mbox{v}_{\phi }\left(r_{\perp }\right)\nonumber \\
&=&2\pi {\hbar \over m}\sum_{\mu }\ m_{\mu }\ \lambda_{\mu }\ 
\frac{\left| \chi_{\mu }\left(r_{\perp},\phi\right)\right|^{2}}
{\rho\left(r_{\perp},\phi\right)}.
\label{crcln}
\end{eqnarray}
These are plotted in Figs. 2-4. For completeness we also give the
velocity curl
\begin{eqnarray}
\nabla\times{\bf v}
&=&\frac{\hat{z}\ \omega_{\perp}}{\rho\left(r_{\perp},\phi\right)} 
\sum_{\mu }\lambda_{\mu }\left(\frac{m_{\mu }}{r_{\perp }\alpha_{\perp }}-
\mbox{v}_{\phi}\sqrt{\frac{m}{\hbar\omega_{\perp }}}\right) 
\left[\chi_{\mu }^{\ast}
\left( \sum_{n_{\mu }}c^{\mu }_{n_{\mu },m_{\mu },0} 
\right.\right. \nonumber \\ && \left.\left. 
\left\{ n_{\mu }\ u_{n_{\mu },m_{\mu },0} 
-\sqrt{\left(n_{\mu }-\mid m_{\mu }\mid\right)
\left(n_{\mu }+\mid m_{\mu }\mid\right)}\ u_{n_{\mu }-2,m_{\mu },0}
\right\}
\right)~+~ c.c \ \right].
\end{eqnarray}
The expression for circulation as in equation(\ref{crcln}) calls for some
discussion. The circulation associated with each of the 
 eigencomponents$\left(\chi_{\mu }\right)$ 
 of the one-particle density-matrix  
 is, of course, quantized as integral multiples of $\frac{h}{m}$ for 
 any closed path, and clearly can have 
  no $r_{\perp }$ dependence for the circular paths chosen
 for our line integral. 
The expression in equation(\ref{crcln}), however, involves an
average over different components $\chi_{\mu }$, and  
can, therefore, assume non-integral values that vary with
 $r_{\perp }$ for the chosen circular path $r_{\perp }$=constant. 
 This reflects the radial variation of the relative
 fractional weights $m_{\mu }\ \lambda_{\mu }\ 
\left| \chi_{\mu }\left(r_{\perp},\phi\right)\right|^{2}$ 
 of the different fractions  
 with $r_{\perp }$.
\section{RESULTS AND DISCUSSION}
\indent
The results and discussions given below refer to the following choice 
of parameters: 
 $a_{\perp }=1.222\mu m$, $\lambda_{z}=\sqrt{8}$, 
 $a_{sc}=1000a_{0}$, where $a_{0}$ is the Bohr radius. Please note that
 the scattering length $a_{sc}$ chosen above is 10 times more than as given
 in \cite{dal96} for $^{87}$Rb. 
Further, we present here the results for N=15 only in the angular
 momentum regime $0\leq \mbox{L}_{z}\leq 3\mbox{N}$. We have, however,
 done the calculations for N=6 and N=10 also.  
 First we note that we
are always in the dilute gas limit here inasmuch as 
$\mbox{N}\left(\frac{a_{sc}}{a_{ho}}\right)^{3}\ll 1$, 
 with $a_{ho}\equiv \left(a_{\perp }^{2}a_{z}\right)^{\frac{1}{3}}
 = a_{\perp }\lambda_{z}^{-\frac{1}{6}}$. Further, we are also in the 
 regime of moderate $\left(a_{sc}=1000a_{0}\right)$ to 
 very weak$\left(a_{sc}=1a_{0}\right)$ interaction strength as
 the parameter $\mbox{N}\frac{a_{sc}}{a_{ho}}$ is of order 1 
 or much much less than 1
 for the two regimes respectively. Moreover for $a_{sc}=1000a_{0}$, 
 the healing length 
 $\xi (\equiv \left(8\pi n a_{sc}\right)^{-\frac{1}{2}}$, with
 $n \sim \frac{\mbox{N}}{a_{ho}^{3}}$, the mean particle density)  
 $<$ the oscillator length
 $a_{ho}\sim $ the size of the system. Thus our study for
 the above choice of paremeters may be
relevant to bulk systems.\\
\indent  
The critical angular velocity $\Omega_{c1}$ for the  
single-vortex state$\left(\mbox{L}_{z}=\mbox{N}\right)$ decreases 
 for increasing value of N. Thus
 for $a_{sc}=1000a_{0}$, it is $0.88487\omega_{\perp }$ and
 $0.84493\omega_{\perp }$ for N=10 and 15 respectively. 
 For a given N, the critical angular velocity $\Omega_{c1}$ is also 
 found to decrease with inceasing scattering length $a_{sc}$.
 Thus for N=15, $\Omega_{{\bf c}1}$(in units of $\omega_{\perp}$)
  are for found to be 
 $.99978$, $.99783$, $0.97901$ and $0.84493$ 
 for $a_{sc}=1a_{0},\ 10a_{0},\ 100a_{0},\ 
1000a_{0}$ respectively. The condensate fraction for a given N and
 corresponding to one of the stable vortex states, however, seems to
 be quite insesitive to a change in the scattering length. Thus
 for L$_{z}$=N=15 state and $a_{sc}=1a_{0},\ 10a_{0},\ 100a_{0},\ 
 1000a_{0}$, the
 condensate fractions$\left(\lambda_{1}\right)$ are found to be 
 $0.87826767,\ 0.87822606,\ 0.87777931,\ 0.87069438$ 
 respectively.\\ 
\indent  
In Table 1, we summarize our results for the ground state
 of the N=15 system in different subspaces of the
 total angular momentum L$_{z}$ in the regime $0\leq \mbox{L}_{z}
 \leq 3\mbox{N}$ for $a_{sc}=1000a_{0}$. 
As the system is rotated, a number of (total) angular momentum
 states are found to be stable at successively higher angular velocities
 $\Omega_{{c}i},\ i=1,2,\cdots $.
 We have also given the density-matrix eigenvalues$\left\{\lambda_{\mu }
 \right\}$ and the corresponding single-particle angular momentum
 quantum number$\left\{m_{\mu }\right\}$ for the largest three
  fractions$\left(\lambda_{1}>\lambda_{2}>\lambda_{3}\right)$. 
  We note that, corresponding to 
 each of the stable ground states of the rotating system,  
  namely, L$_{z}=15,\ 26,\ 30,\ 33,\ 35,\ 36,\ 45,\cdots $ for N=15, 
 the second largest fraction is found to be non-rotating, {\em
 i.e.} $m_{2}=0$. The single-particle angular momentum   
 quantum number $m_{1}$, corresponding to the largest fraction, 
  is taken to be the vorticity of the condensate.\\
\indent  
 Figure 1 gives the L$_{z}-\Omega$ phase diagram, or the stability
 line. One can readily identify the successive critical angular 
velocities $\left\{ \Omega_{{\bf c}i},\ i=0,1,2,\cdots \right\}$ 
 for the transition from one
stable state to another stable state. Several points are 
to be noted here. The non-integral values for the angular
momentum per particle clearly indicates the incomplete 
 condensation, i.e., the fact that more than one eigenvalues
 $\left\{\lambda_{\mu }\right\}$ of the reduced density-matrix are non-zero.
 This is, however, not to be taken as the fragmented 
 condensate\cite{noz82}, but merely as depletion
 of the condensate/superfluid fraction due to the interaction and 
 the rotation.   
 Also, we have observed that the critical angular velocity 
 $\Omega_{{\bf c}i}$ decreases
 monotonically with increasing repulsive interaction and the particle
 number. This can be
 readily understood physically from the expression in equation(\ref{4p4}) 
 for the critical
 angular velocity $\Omega_{{\bf c}i}\left(\mbox{L}_{z}\right)$ 
 if one remembers that the non-rotating (L$_{z}=0$ angular
 momentum) state is more compact and, therefore, has its energy
 raised (because of the repulsive interaction) relative to the 
 expanded higher angular momentum states. More importantly, however,
 our critical angular velocity $\Omega_{{\bf c}i}$ for a given N, is 
 systematically greater
 than the non-variational value based on the Yrast-like state  
 \cite{rokh99,bmot99,gfb99,wg00}. 
 We expect this to be due to our more accurate and, 
 therefore, lower variational ground state energies.\\
\indent
In Fig.(2a) we have plotted the total
(the condensate+the non-rotating component+the 
 above-the-condensate fraction)
 density profile $\rho\left(r_{\perp }\right)$ 
 for the L$_{z}$=N single-vortex angular 
 momentum state of the system. We have also plotted separately 
 the density profiles for the condensate fraction 
$\lambda_{1}\left|\chi_{1}\left({\bf r}\right)\right|^{2}$
corresponding to the largest eigenvalue $\lambda_{1}
\left(=0.87069\right)$ in Fig.(2b),
 the non-rotating component 
$\lambda_{2}\left|\chi_{2}\left({\bf r}\right)\right|^{2}$  
 corresponding to the second largest eigenvalue $\lambda_{2}
 \left(=0.06576\right)$ in Fig.(2c),  
 and the above-the condensate fraction in Fig.(2d) corresponding
 to the remaining
 components of the one-particle reduced density matrix: 
\begin{eqnarray*} 
\rho\left({\bf r},{\bf r}^{\prime }\right)~=~
\underbrace{\lambda_{1}\chi_{1}^{\ast }\left({\bf r}\right)~ 
\chi_{1}({\bf r}^{\prime })}_{
\stackrel{\textstyle the~condensate}{\textstyle with~m_{1}=1} 
                             }
 ~+~\underbrace{\lambda_{2}\chi_{2}^{\ast }({\bf r})
 \chi_{2}({\bf r}^{\prime })}_{
 \stackrel{\textstyle the~non\mbox{-}rotating~com}{\textstyle 
        \mbox{-}ponent~with~m_{2}=0} 
                              }
~+~\underbrace{\sum_{\mu =3}~\lambda_{\mu } \chi_{\mu }^{\ast }
 ({\bf r})~\chi_{\mu }({\bf r}^{\prime })}_{
 \stackrel{\textstyle the~above\mbox{-}the}
 {\textstyle \mbox{-}condensate}}
\end{eqnarray*}
where, as we have  earlier, $m_{\mu }$ is the single-particle
 angular momentum quantum number associated with the $\mu $th 
 component of the density-matrix. As we go to higher $\Omega $,
 the condensate depletes$-$with the non-rotating component and the
 above-the-condensate fractions becoming more pronounced(as can also
be seen from Table 1).\\   
\indent
Figure 3 depicts the velocity profile for the total system(3a), 
 for the condensate(3b), for the non-rotating  
 component(3c), and for the above-the-condensate fractions(3d) respectively.\\
\indent
 Figure 4 depicts the circulation profile for circular paths 
 of radii $r_{\perp }$ in the
 x-y plane: (4a) for the total system, (4b) for the condensate fraction, 
 (4c) for the non-rotating 
 fraction, and (4d) for the above-the-condensate fraction respectively. 
 As can 
 readily be seen, the condensate fraction has an integral circulation
  and is constant over space, while the total as well as the 
 above-the-condensate 
  fraction has non-integral circulation and varies with  
 $r_{\perp }$ as explained earlier below equation(\ref{crcln}).\\
\indent
In conclusion, our many-body variational calculation for an
admittedly finite number $\left(\mbox{N}\leq 15\right)$ of particles 
 explicitly resolves the structure of the rotating BEC at T=0 in terms of   
 the various components of the reduced one-particle density-matrix. 
 Further work is in progress to consider spin-1 Bose particles, 
 where we expect qualitative changes in the structure of the
 rotating BEC. In particular,  we expect a dramatic increase in the critical 
 angular velocities because of the spin-polarization drag effect inasmuch as
 the angular momentum may be absorbed preferentially in the spin rather than
 the orbital degrees of freedom without costing kinetic energy. This
 could lead to spin polarization due to rotation of the 
 trap\cite{jaynes,nank}. We expect this effect 
 to be more pronounced here than in a classical rotating system.

\section*{Acknowledgment} 
\noindent
One of us(MAHA) would like to thank Prof. 
R. Nityananda and Prof. R. Srinivasan for useful discussions. 

\newpage
\thispagestyle{empty}
\noindent   
\underline{\bf Table caption}\\
Table 1. Gives the largest three condensate fractions 
$\lambda_{1} > \lambda_{2} > \lambda_{3} $ and the 
corresponding single-particle angular momentum eigenvalues 
$m_{1}$, $m_{2}$ and $m_{3}$ in the one-particle reduced density-matrix 
for the ground state of the rotating BEC, in given subspaces of 
total angular momentum L$_{z}$. 
Also given are the ground state energies 
 E$_{0}^{lab}\left(\mbox{L}_{z} \right)$ in the laboratory frame
 and the critical angular velocities $\Omega_{{\bf c}i}$ 
corresponding to the stable ground states of the rotating
system in the angular momentum regime 
$0\leq \mbox{L}_{z}\leq 3\mbox{N}$.\\  

\noindent
\underline{\bf Figure captions}\\
\noindent
FIG.1. Gives the plot (solid line) for 
the total angular momentum L$_{z}$ in units of 
 $\hbar $ 
 {\em versus} 
the critical rotational 
 velocity $\Omega_{{\bf c}i}$(in units of $\omega_{\perp })$
for the rotating BEC. Dashed line is for the 
 non-interacting case, included here for reference.\\

\noindent
FIG.2. Depicts the normalized 
 particle-density in units of $a_{\perp }^{-2}$ for, 
 (a) the total system, (b) the condensate fraction, (c) the
 non-rotating component, and (d) the above-the-condensate fraction, in the
 x-y plane. Distances are in units of $a_{\perp }$.\\ 
  
\noindent
FIG.3. Depicts the azimuthal velocity profile 
 $v_{\phi }\left(r_{\perp}\right)$ 
  in units of $\sqrt{\frac{\hbar\omega_{\perp }}{m}}$ for,  
 (a) the total system, (b) the condensate fraction,  (c) the
 non-rotating component, and (d) the above-the-condensate fraction, in the
 x-y plane. Distances are in units of $a_{\perp }$.\\ 
  
\noindent
FIG.4. Depicts the circulation $ \kappa(r_{\perp }) \equiv \oint {\bf v}\cdot d{\bf l} $ profile 
 in units of $\frac{h}{m}$ along circular paths of radius  
 $r_{\perp }$ in the x-y plane for,  
 (a) the total system, (b) the condensate fraction, (c) the
 non-rotating component,  
 and (d) the above-the-condensate fraction. 
 Distances are in units of $a_{\perp }$.\\ 
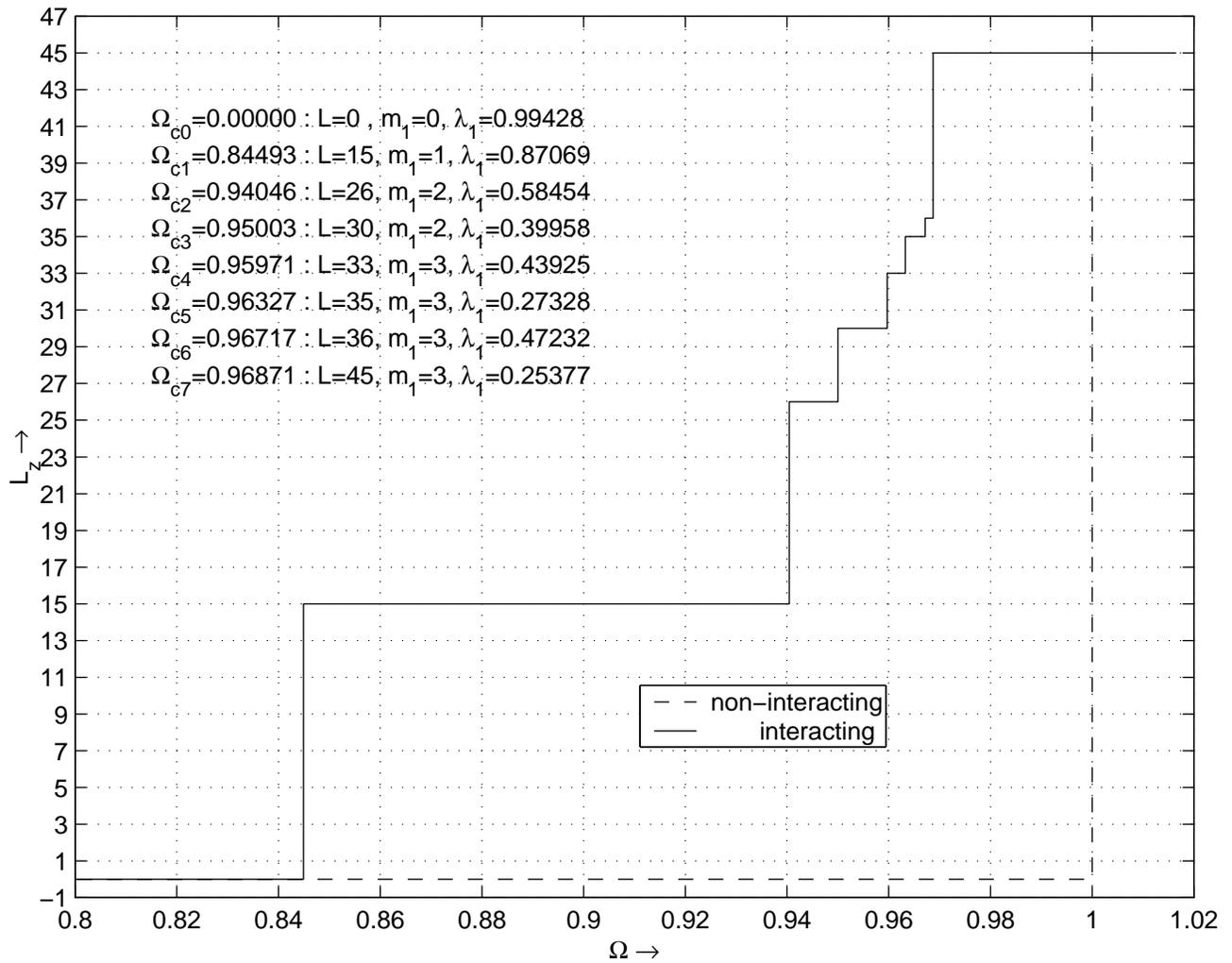
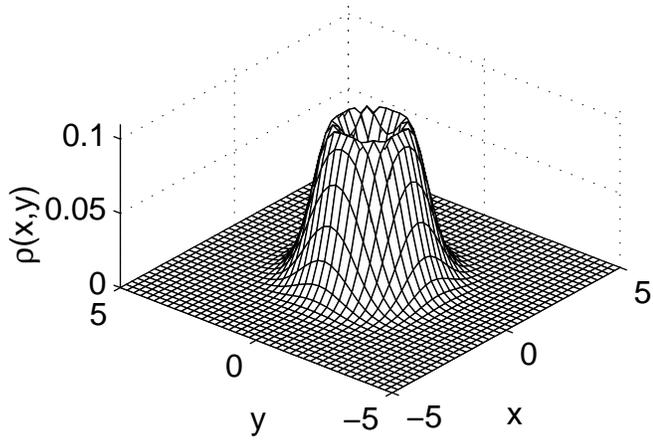
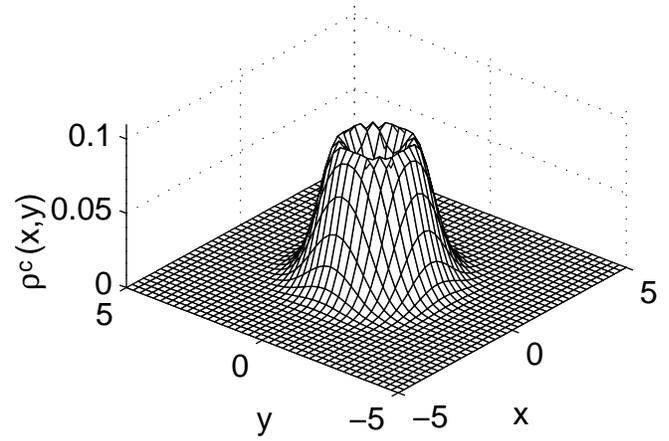
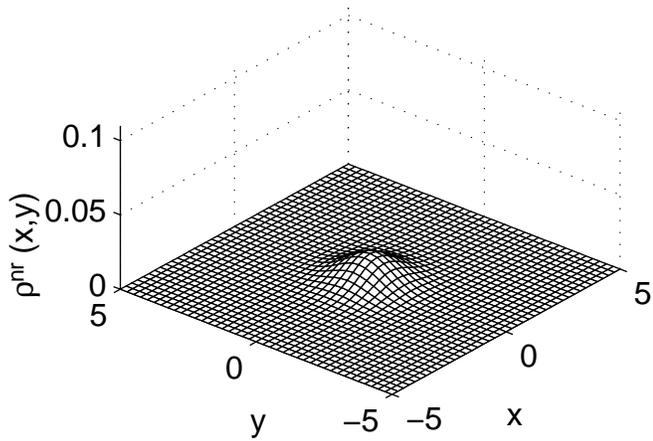
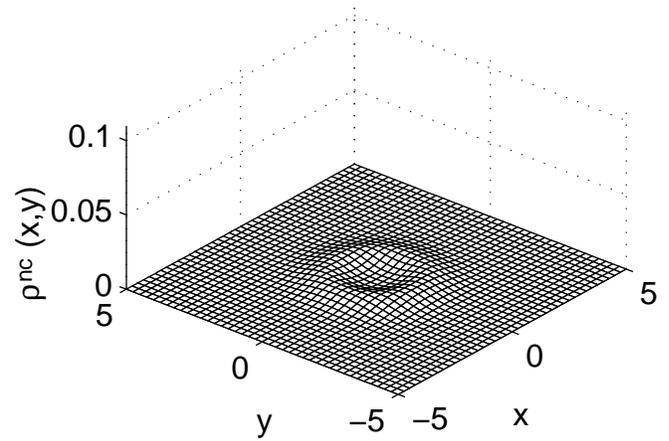
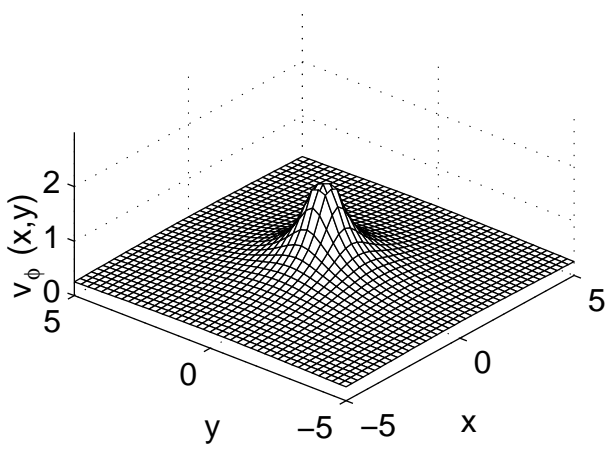
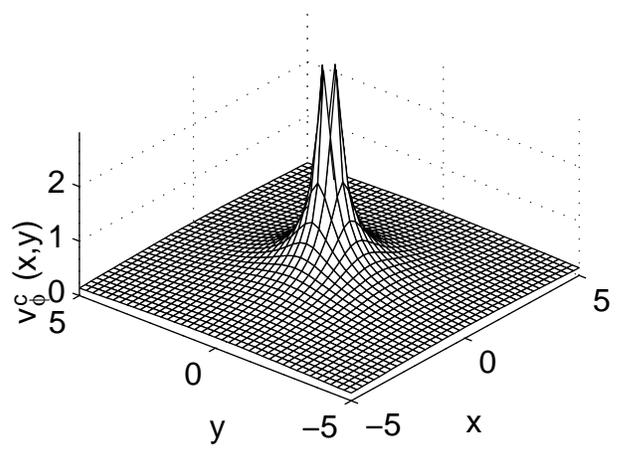
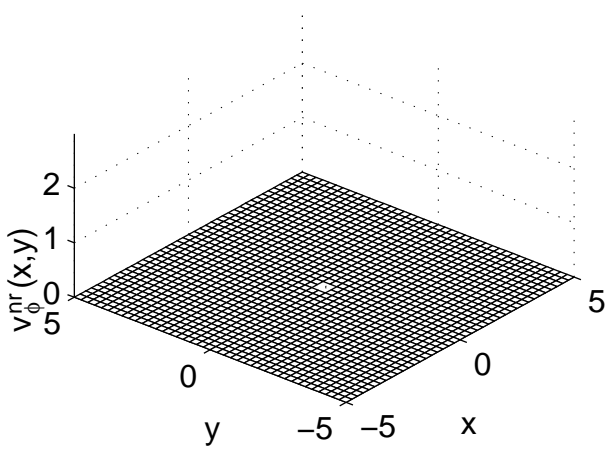
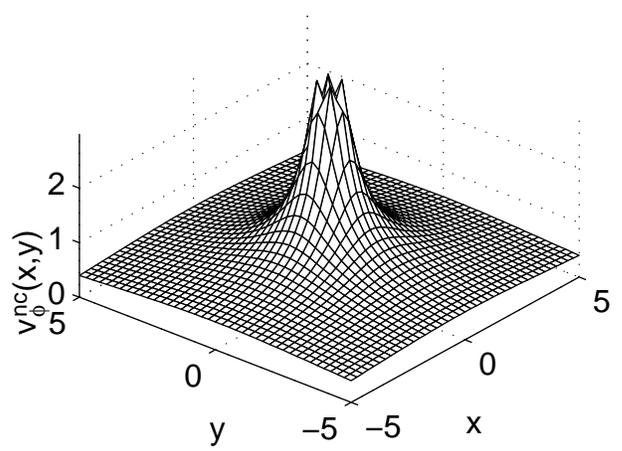
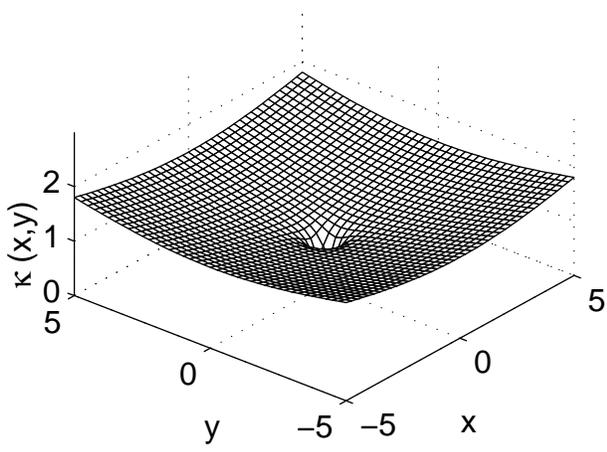
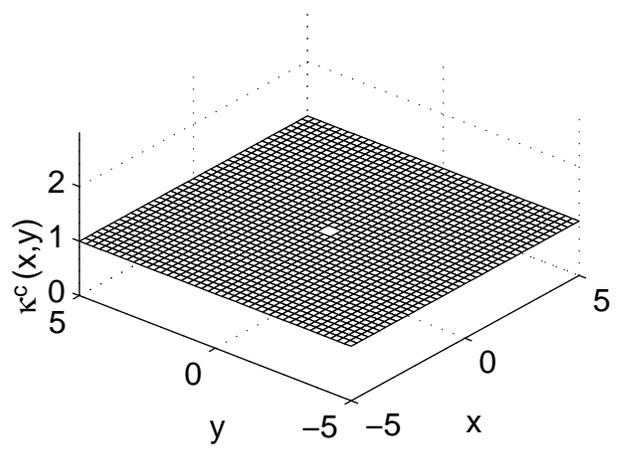
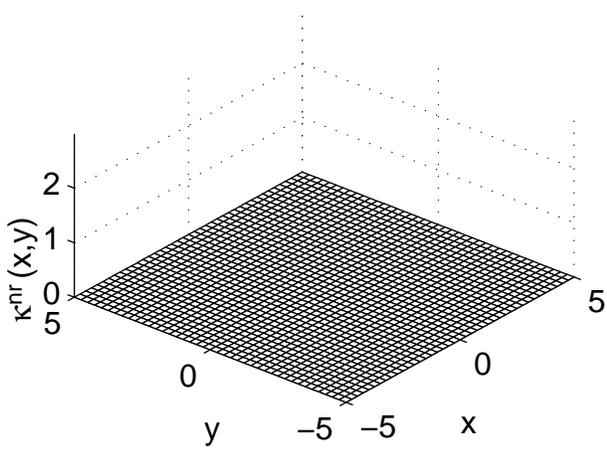
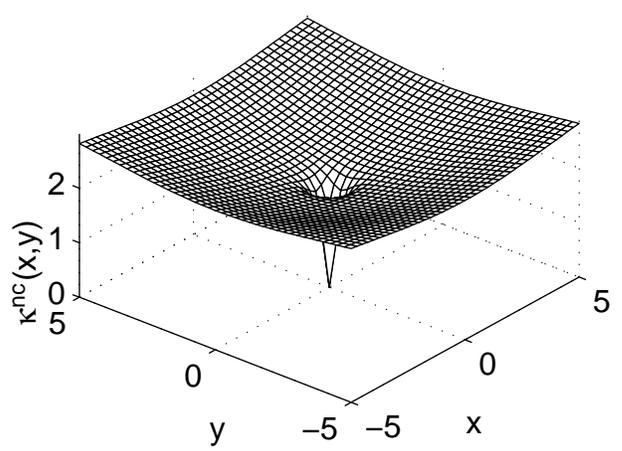
\end{document}